\documentclass[a4paper,fleqn]{cas-dc}

\usepackage[authoryear,longnamesfirst]{natbib}

\usepackage{booktabs} 
\usepackage{multirow}
\usepackage{graphicx}
\usepackage{subcaption}
\usepackage{xcolor}
\usepackage{tabularx}
\usepackage{enumitem}
\usepackage{url}
\usepackage{algorithm}
\usepackage{algpseudocode}

\def\tsc#1{\csdef{#1}{\textsc{\lowercase{#1}}\xspace}}
\tsc{WGM}
\tsc{QE}
\tsc{EP}
\tsc{PMS}
\tsc{BEC}
\tsc{DE}

\begin{document}
\let\WriteBookmarks\relax
\def\floatpagepagefraction{1}
\def\textpagefraction{.001}

\shorttitle{Modernizing SIENA with Targeted Deep Learning Integration}

\shortauthors{R. Raciti et al.}

\title [mode = title]{Reinforcing the Weakest Links: Modernizing SIENA with Targeted Deep Learning Integration}


\author[1]{Riccardo Raciti}[orcid=0009-0003-2389-4879]
\cormark[1]
\ead{riccardo.raciti@phd.unict.it}
\credit{Conceptualization, Data curation, Formal analysis, Investigation, Methodology, Software, Validation, Visualization, Writing – original draft}

\author[1]{Lemuel Puglisi}
\credit{Software, Writing – review and editing}

\author[1]{Francesco Guarnera}
\credit{Supervision, Writing – review and editing}

\author[2]{Daniele Ravì}
\credit{Supervision, Writing – review and editing}

\author{for the Alzheimer's Disease Neuroimaging Initiative}
\fnmark[1]

\author[1]{Sebastiano Battiato}
\credit{Conceptualization, Methodology, Project administration, Supervision, Writing – review and editing}


\affiliation[1]{organization={Dept. of Math and Computer Science, University of Catania},
                addressline={Viale Andrea Doria, 6}, 
                city={Catania},
                country={Italy}}

\affiliation[2]{organization={MIFT Department, University of Messina}, 
                addressline={Viale Ferdinando Stagno d'Alcontres 31},
                city={Messina},
                country={Italy}}

\cortext[cor1]{Corresponding author}

\fntext[fn1]{
Data used in preparation of this article were obtained from the Alzheimer's Disease Neuroimaging Initiative (ADNI) database (adni.loni.usc.edu). As such, the investigators within the ADNI contributed to the design and implementation of ADNI and/or provided data but did not participate in the analysis or writing of this report. A complete listing of ADNI investigators can be found at: \url{http://adni.loni.usc.edu/wp-content/uploads/how\_to\_apply/ADNI\_Acknowledgement\_List.pdf} 
}


\begin{abstract}
Percentage Brain Volume Change (PBVC) derived from Magnetic Resonance Imaging (MRI) is a widely used biomarker of brain atrophy, with SIENA among the most established methods for its estimation. However, SIENA relies on classical image processing steps, particularly skull stripping and tissue segmentation, whose failures can propagate through the pipeline and bias atrophy estimates. In this work, we examine whether targeted deep learning substitutions can improve SIENA while preserving its established and interpretable framework. To this end, we integrate SynthStrip and SynthSeg into SIENA and evaluate three pipeline variants on the ADNI and PPMI longitudinal cohorts. Performance is assessed using three complementary criteria: correlation with longitudinal clinical and structural decline, scan-order consistency, and end-to-end runtime. Replacing the skull-stripping module yields the most consistent gains: in ADNI, it substantially strengthens associations between PBVC and multiple measures of disease progression relative to the standard SIENA pipeline, while across both datasets it markedly improves robustness under scan reversal. The fully integrated pipeline achieves the strongest scan-order consistency, reducing the error by up to 99.1\%. In addition, GPU-enabled variants reduce execution time by up to 46\% while maintaining CPU runtimes comparable to standard SIENA. Overall, these findings show that deep learning can meaningfully strengthen established longitudinal atrophy pipelines when used to reinforce their weakest image processing steps. More broadly, this study highlights the value of modularly modernizing clinically trusted neuroimaging tools without sacrificing their interpretability. Code is publicly available at \url{https://github.com/Raciti/Enhanced-SIENA.git}.
\end{abstract}


\maketitle

\section{Introduction}
Quantifying disease progression is a central challenge in neurodegenerative disorders such as Alzheimer's disease (AD) and Parkinson's disease (PD). As the prevalence of these conditions increases with aging populations~\citep{feigin2017global}, sensitive biomarkers of neurodegeneration are essential for evaluating disease-modifying therapies and detecting clinically meaningful effects in clinical trials. Traditionally, the progression of neurodegenerative diseases has been assessed using clinician-rated scales and cognitive or motor tests~\citep{cockrell2002mini,rudick1999use,rosen1984new,hughes1982new,pfeffer1982measurement,nasreddine2005montreal}. Although widely used, these measures are influenced by rater and test-retest variability and often require long follow-up to detect change~\citep{kadirvelu2023wearable}. They may also be insensitive to early neurobiological progression, motivating quantitative biomarkers that provide objective and reproducible measures of neurodegeneration~\citep{coley2011suitability,mcdougall2021psychometric,brem2023digital}.

Structural Magnetic Resonance Imaging (MRI) is well suited for longitudinal assessment of neurodegeneration because it provides non-invasive measures of brain structure that can be acquired repeatedly over time. In this context, a central MRI-derived marker is brain atrophy, typically quantified as global or regional volume loss. For longitudinal studies, atrophy is often summarized as Percentage Brain Volume Change (PBVC), defined as the percentage change in brain volume between two scans acquired at different time points. PBVC has been reported to correlate with clinical progression and cognitive decline in disorders such as AD~\citep{henneman2009hippocampal} and PD~\citep{mak2017longitudinal}.
Estimating PBVC robustly from paired MRI scans remains challenging, particularly in the presence of large anatomical changes and scan-to-scan variability arising from acquisition and preprocessing differences. A widely used approach is SIENA~\citep{siena1,siena2}, which estimates PBVC by analyzing the displacement of brain boundaries between two MRI scans after co-registration and tissue segmentation. Although extensively validated, SIENA relies on classical image processing algorithms implemented in the FSL toolbox~\citep{smith2004advances}, including the Brain Extraction Tool (BET2)~\citep{jenkinson2005bet2} for skull stripping and FAST~\citep{Fast} for tissue segmentation. These algorithms rely on intensity-based heuristics and deformable surface models, and their performance may degrade in the presence of severe neurodegeneration or imaging artifacts~\citep{nakamura2018improving,klauschen2009evaluation}. In particular, errors in brain extraction can trigger a cascading failure, propagating to downstream registration and tissue-segmentation steps and ultimately biasing PBVC estimates.

Recent advances in Deep Learning (DL) make it possible to estimate the PBVC directly from paired MRI scans in an end-to-end manner~\citep{raciti2024efficient}. Although these methods can be effective, they are typically less transparent and explainable than SIENA. An alternative, more interpretable strategy is to use DL to improve individual components of established pipelines. DL approaches perform well in key image processing tasks, including tissue segmentation~\citep{billot2023synthseg} and brain extraction~\citep{hoopes2022synthstrip}. However, it is not yet clear how substituting SIENA’s corresponding image processing steps with DL-based alternatives influences SIENA-derived PBVC, and this question has not been systematically studied. To address this gap, we revisit the SIENA framework and evaluate whether DL-based image processing can improve the robustness and clinical sensitivity of longitudinal brain atrophy estimation. Specifically, this paper makes the following contributions:\newline

\begin{itemize}
\item We investigate the integration of DL–based image processing within the SIENA pipeline by replacing its brain extraction and tissue segmentation steps with SynthStrip~\citep{hoopes2022synthstrip} and SynthSeg~\citep{billot2023synthseg}, yielding three SIENA pipeline variants.
\item We evaluate these variants against the original SIENA pipeline on two longitudinal cohorts of AD and PD patients, assessing their impact on PBVC estimation using three complementary criteria: (i) association with disease progression, (ii) directional symmetry under scan order reversal, and (iii) computational efficiency.
\end{itemize}

Our experiments indicate that updating SIENA’s image processing steps with DL–based components can improve both the clinical sensitivity and the robustness of PBVC estimates, with particularly consistent gains when the skull-stripping step is replaced. We also observe improved directional symmetry when combining DL–based skull stripping and segmentation, while maintaining comparable end-to-end runtime. Together, these results suggest that targeted modernizations can strengthen established neuroimaging pipelines without altering SIENA’s core methodology.

\section{Related work}\label{sec:sota}
In this section, we review prior work on longitudinal MRI-based atrophy assessment, discuss the main limitations of SIENA, and summarize recent DL approaches for robust brain MRI image processing.

\subsection{Longitudinal atrophy assessments}
Quantifying structural brain changes from longitudinal MRI is a key task in neuroimaging studies of neurodegenerative diseases. Early work introduced the Boundary Shift Integral (BSI)~\citep{freeborough1997boundary}, which estimates volume change by integrating intensity differences across brain boundaries after registration of longitudinal scans. However, it initially required manual intervention at the brain-extraction stage~\citep{smith2007longitudinal}. Subsequently, SIENA~\citep{siena1,siena2} was proposed as a fully automated approach to estimate PBVC from paired MRI scans. SIENA measures atrophy from the displacement of brain edges following registration and tissue segmentation, and has become one of the most widely adopted pipelines for longitudinal whole-brain atrophy estimation. Previous studies have shown that SIENA and BSI provide broadly comparable atrophy estimates and similar sensitivity for detecting disease-related change~\citep{smith2007longitudinal}. More recently, DL approaches such as DeepBVC~\citep{zhan2023learning} and Efficient Atrophy Mapping (EAM)~\citep{raciti2024efficient} have attempted to directly predict SIENA-derived brain boundary shifts using 3D convolutional neural networks. Although this end-to-end formulation is appealing, it relies on SIENA-generated targets, which are inherently noisy and therefore provide an imperfect supervisory signal. Furthermore, directly predicting boundary shifts reduces the transparency of classical pipelines such as SIENA, where each processing stage and its intermediate outputs can be explicitly inspected. BrainLossNet~\citep{opfer2024brainlossnet} instead estimates PBVC using deformation-driven mask warping. However, its raw output is affected by scanner-related distortions and is not expressed on the PBVC scale; therefore, the method first applies distortion correction and then rescales the corrected signal using SIENA-derived PBVC estimates from the training data, making the final prediction dependent on SIENA outputs. These limitations motivate a modular strategy in which DL enhances individual components within established longitudinal atrophy estimation frameworks.

\begin{figure*}[pos=t]
     \centering
     \begin{subfigure}[b]{0.48\linewidth}
         \centering
         \includegraphics[width=\linewidth]{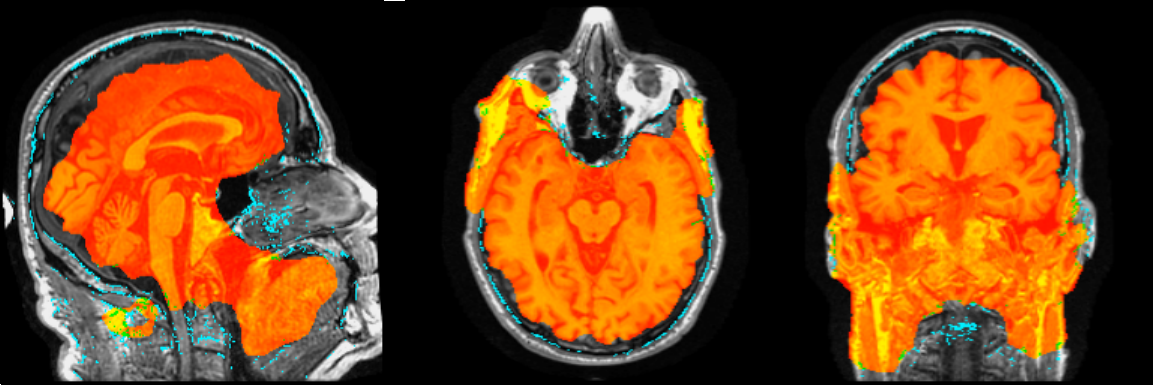}
         \caption{BET2 Skull Stripping}
         \label{fig:skull_stripping_edg_comparison_BET}
     \end{subfigure}
     \hfill
     \begin{subfigure}[b]{0.48\linewidth}
        \centering
         \includegraphics[width=\linewidth]{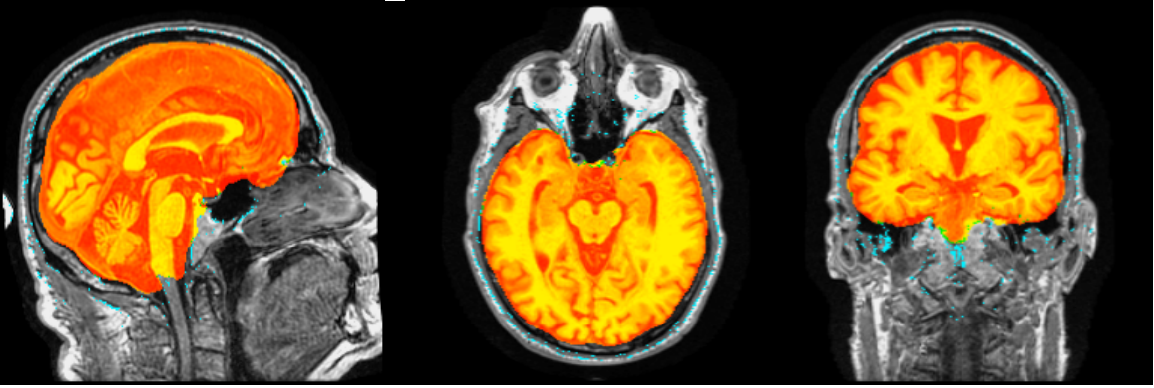}
         \caption{SynthStrip Skull Stripping}
         \label{fig:skull_stripping_edg_comparison_SS}
     \end{subfigure}

     \vspace{0.5cm}

     \begin{subfigure}[b]{0.48\linewidth}
         \centering
         \includegraphics[width=\linewidth]{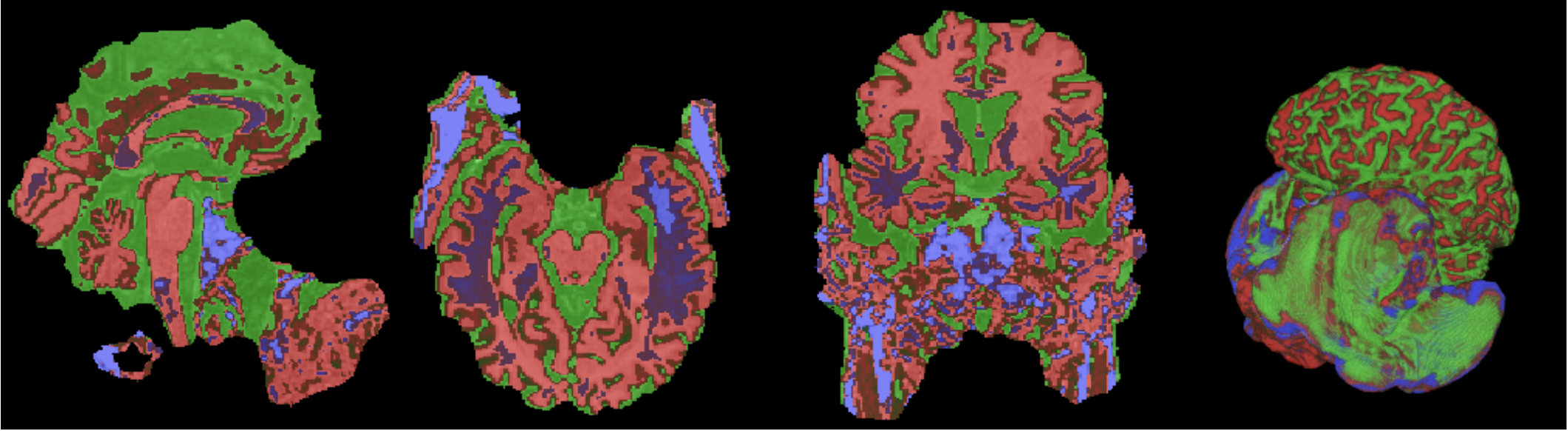}
         \caption{FAST Segmentation on BET2 result}
         \label{fig:seg_comparison_FAST}
     \end{subfigure}
     \hfill
     \begin{subfigure}[b]{0.48\linewidth}
        \centering
         \includegraphics[width=\linewidth]{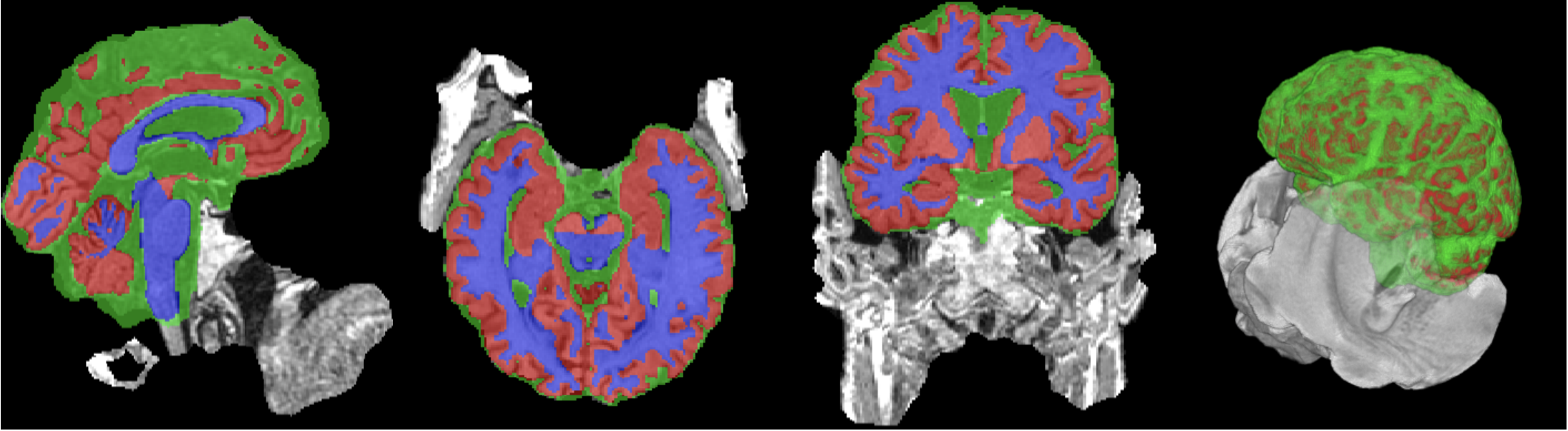}
         \caption{SynthSeg Segmentation on BET2 result}
         \label{fig:seg_comparison_SS}
     \end{subfigure}

    \caption{Qualitative comparison of traditional and deep-learning frameworks for brain extraction and segmentation. (a) Brain extraction results using BET2. (b) Brain extraction results using SynthStrip. (c) Tissue segmentation results using FAST, performed on the BET2 extraction output. (d) Tissue segmentation results using SynthSeg, performed on the BET2 extraction output (labels aggregated for comparison as described in Section~\ref{sec:synthseg-integ}). Compared to BET2 (a), SynthStrip (b) provides more anatomically consistent and robust exclusion of non-brain structures. Furthermore, SynthSeg (d) demonstrates greater robustness than FAST (c) when processing a suboptimal initial brain extraction.}
    \label{fig:tool_comparison}
\end{figure*}

\subsection{Critical limitations of SIENA}
Despite its widespread adoption, the accuracy of SIENA depends heavily on the robustness of its image processing steps. In particular, the pipeline relies on classical algorithms from the FSL toolbox~\citep{smith2004advances}, including BET2~\citep{jenkinson2005bet2} for skull stripping and FAST~\citep{Fast} for tissue segmentation. Several studies have shown that inaccuracies in the initial brain extraction stage can substantially affect downstream PBVC estimates, as the quality of the brain mask directly influences subsequent registration and boundary detection steps. For example, small variations in the BET2 fractional intensity parameter can lead to significant differences in measured atrophy rates~\citep{nakamura2018improving}. Furthermore, BET2 has been shown to be sensitive to signal inhomogeneities, motion artifacts, and anatomical abnormalities~\citep{ou2018field,mohapatra2023sam}, with these effects sometimes causing the inclusion of non-brain tissue or the inadvertent removal of cortical structures (see Figure~\ref{fig:skull_stripping_edg_comparison_BET}). These inaccuracies can propagate through the pipeline and negatively affect the reliability of longitudinal atrophy estimates~\citep{tudorascu2016reproducibility}. Such issues are particularly problematic in neurodegenerative populations, where severe atrophy and structural abnormalities challenge traditional intensity-based extraction methods~\citep{nakamura2018improving,eskildsen2012beast}. Although FAST can also be affected by pathology and imaging heterogeneity~\citep{klauschen2009evaluation}, its limitations are often most consequential when upstream skull-stripping is imperfect (see Figure~\ref{fig:seg_comparison_FAST}), as FAST requires optimal brain–non-brain separation to function well. This makes robust brain extraction, as well as the investigation of more independent segmentation tools, particularly important for improving SIENA in challenging clinical datasets.

\subsection{Robust DL-based image processing}
Recent advances in DL have led to substantial improvements in medical image processing tasks~\citep{hoopes2022synthstrip, billot2023synthseg, rondinella2023boosting}. A particularly important development has been the emergence of frameworks trained through domain randomization~\citep{hoopes2022synthstrip,billot2023synthseg,hoffmann2021synthmorph,puglisi2024synthba}, in which fully synthetic images are generated on the fly from anatomical label maps while contrast, resolution, noise, and artifacts are stochastically varied during training. By exposing models to highly diverse and even atypical image appearances, this strategy promotes strong generalizability across acquisition protocols and common real-world imaging challenges. Several tools have applied this concept to key brain MRI image processing steps, including SynthStrip~\citep{hoopes2022synthstrip} for brain extraction (see Figure~\ref{fig:skull_stripping_edg_comparison_SS}) and SynthSeg~\citep{billot2023synthseg} for semantic tissue segmentation (see Figure~\ref{fig:seg_comparison_SS}). Together, these approaches have demonstrated that domain-randomized DL models can substantially improve the robustness and generalizability of core MRI image processing tasks compared with classical methods. However, their potential impact on established longitudinal atrophy pipelines such as SIENA has not yet been systematically evaluated.

\section{Background: the SIENA pipeline}\label{sec:background}
SIENA estimates longitudinal brain atrophy from paired MRI scans through a sequence of image-processing steps implemented in FSL~\citep{smith2004advances}. The high-level steps of the pipeline are summarized below.
\begin{itemize}
    \item \textbf{Brain and Skull extraction.} SIENA first applies BET2~\citep{jenkinson2005bet2} to each scan to obtain binary masks for the brain and skull.
    \item \textbf{Symmetric skull-constrained registration.} The two time points are then affinely registered via FLIRT~\citep{flirt2} using both brain and skull masks. The skull serves as a relatively stable anatomical reference, preventing the inadvertent normalization of longitudinal atrophy. The forward and backward transforms are combined to map both scans into a common halfway space, reducing directional bias and balancing interpolation effects.
    \item \textbf{Boundary identification.} FAST~\citep{Fast} is then used on the aligned images to segment each scan into White Matter (WM), Gray Matter (GM), and Cerebrospinal Fluid (CSF). These segmentations are used to identify voxels near the brain boundary.
    \item \textbf{Boundary shift estimation.} For each boundary voxel, SIENA estimates the local surface normal and samples image intensities along this direction in both scans. The displacement that best aligns the two resulting intensity profiles is interpreted as the local boundary motion. Aggregating these displacements over the brain surface yields an estimate of PBVC.
    \item \textbf{Symmetric averaging.} The previous step is performed twice to estimate the PBVC in both the forward and backward directions. The final PBVC estimate is obtained by averaging these two measurements to provide a more robust and unbiased estimate.
\end{itemize}

\section{Method}
\begin{figure*}[pos=t]
    \centering
    \includegraphics[width=\linewidth]{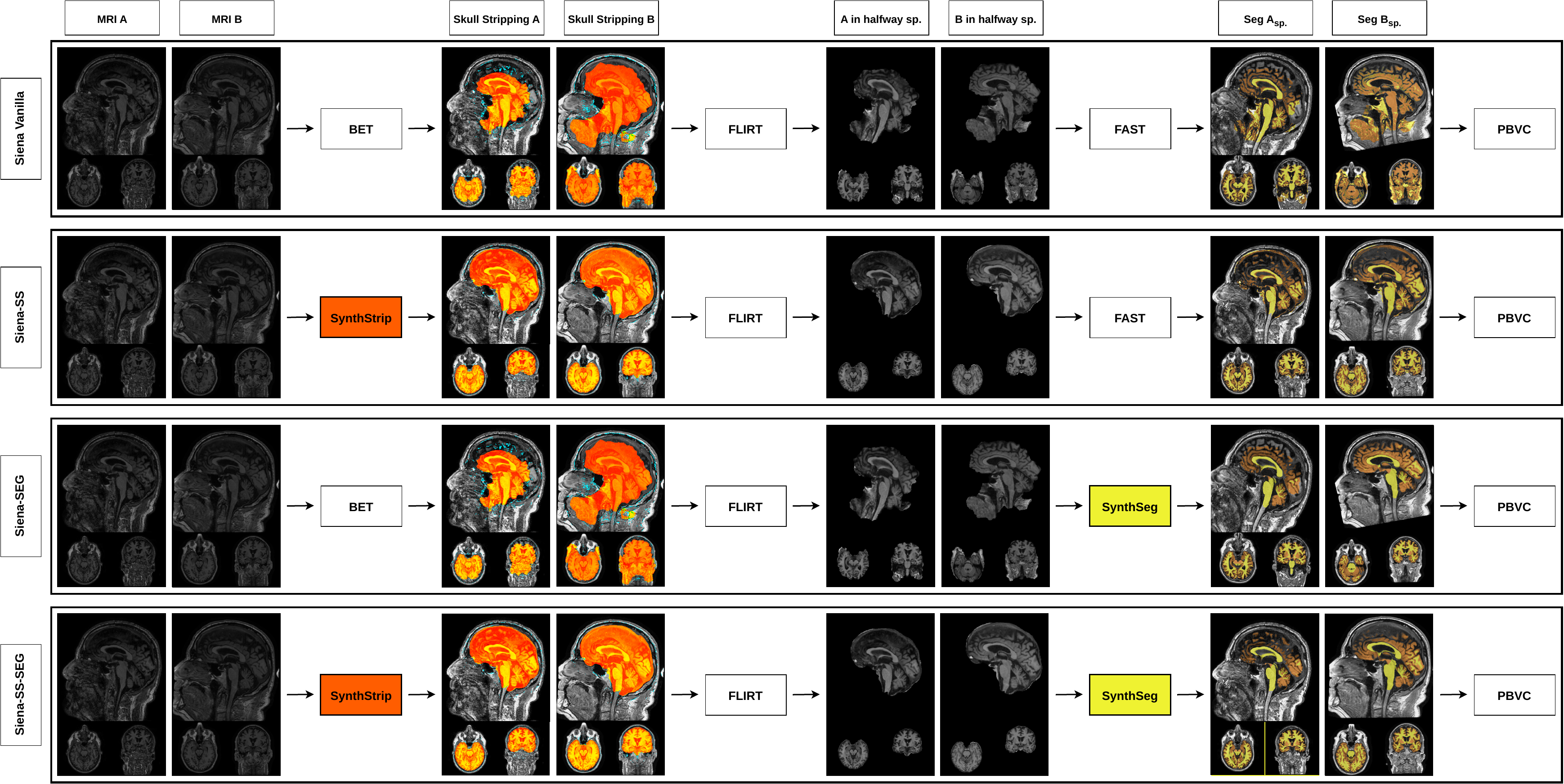}
    \caption{Overview of the four evaluated pipelines. The rows represent \texttt{SIENA Vanilla} (baseline), \texttt{SIENA-SS} (modified skull stripping), \texttt{SIENA-SEG} (modified segmentation) and \texttt{SIENA-SS-SEG} (fully integrated) workflows. Input data consists of baseline and follow-up T1-weighted MRI scans (MRI A and MRI B). The skull stripping step shows the brain extraction results, with orange overlays indicating the segmented brain volume and cyan contours marking the skull-mask boundaries. The FLIRT step shows the affinely co-registered scans in a common halfway space. The segmentation step depicts tissue classification into CSF, GM, and WM using FAST or SynthSeg. All workflows conclude with the estimation of PBVC.}
    \label{fig:pipeline}
\end{figure*}

We evaluate whether replacing selected image processing steps components of the SIENA pipeline with DL alternatives improves the robustness of longitudinal atrophy estimation. The proposed modifications affect two stages: brain extraction and tissue segmentation. Specifically, we integrate SynthStrip for skull stripping and SynthSeg for tissue segmentation, while preserving the remainder of the SIENA pipeline. Figure~\ref{fig:pipeline} summarizes the resulting processing workflows.

\subsection{SynthStrip integration}
As discussed in Section~\ref{sec:background}, SIENA requires both a brain mask and a skull mask to constrain the affine registration between time points. SynthStrip provides a brain mask but does not explicitly estimate a skull mask. To maintain compatibility with SIENA, we implement a procedure to derive an approximate skull mask from the SynthStrip brain segmentation. Specifically, we first smooth the binary brain mask using a Gaussian kernel ($\sigma = 1.0$). We then approximate surface normals from the gradient of the smoothed mask and cast rays outward from boundary voxels along the estimated normal directions up to a maximum distance of 30\,mm. Along each ray, we detect the inner skull boundary using the BET2 intensity-gradient heuristic~\citep{jenkinson2005bet2}. We aggregate the detected points to construct a skull mask, which we use in SIENA's skull-constrained registration stage.

\subsection{SynthSeg integration}\label{sec:synthseg-integ}
The output labels produced by SynthSeg correspond to anatomical structures\footnote{The full list of SynthSeg anatomical labels is available at: \url{https://surfer.nmr.mgh.harvard.edu/fswiki/SynthSeg}} rather than the three tissue classes required by SIENA. To ensure compatibility with the pipeline, we merge these labels into CSF, GM, and WM. Specifically, CSF includes lateral, inferior lateral, third, and fourth ventricles, as well as outer. GM includes cortical and subcortical gray-matter structures, such as the cerebral cortex, thalamus, caudate, putamen, pallidum, hippocampus, amygdala, cerebellar cortex, accumbens, and ventral diencephalon. WM includes cerebral WM, cerebellar WM, and the brainstem. We then pass the resulting three-class segmentation to SIENA for boundary voxel identification.

\subsection{Pipeline variants}
We investigate three pipeline variants, defined by the combination of skull-stripping method (BET2 or SynthStrip) and tissue-segmentation method (FAST or SynthSeg): \texttt{SIENA-SS} (SynthStrip + FAST), \texttt{SIENA-SEG} (BET2 + SynthSeg), and \texttt{SIENA-SS-SEG} (SynthStrip + SynthSeg). The original pipeline is referred to as \texttt{SIENA Vanilla} (BET2 + FAST). All variants share the same SIENA stages for registration, boundary identification, and PBVC estimation. Figure~\ref{fig:pipeline} illustrates these four pipelines.

\subsection{Implementation details}\label{sec:impdetails}
We run all experiments on a workstation equipped with an NVIDIA A100 GPU and an Intel Xeon E5-2650 v4 CPU. We use the SIENA implementation distributed with FSL (v6.0.7.17). We execute SynthStrip and SynthSeg (v2) via FreeSurfer (v7.4.1) using the default command-line interface; for SynthSeg, we enable the \texttt{robust} option. We make the implementation available at \url{https://github.com/Raciti/Enhanced-SIENA.git}.

\section{Experimental settings}\label{sec:experiment}
In this section, we first describe the datasets along with their associated clinical and imaging measures. Next, we examine the relationship between each pipeline’s estimated PBVC and longitudinal changes in cognitive, motor, and structural measures. We then assess pipeline consistency by analyzing how PBVC estimates are affected when the order of MRI scans is reversed. Finally, we compare the pipelines in terms of execution time.

\subsection{Dataset}\label{sec:dataset}
We evaluate all four pipelines using longitudinal T1-weighted brain MRI data from two multi-site cohorts: Alzheimer's Disease Neuroimaging Initiative (ADNI)~\citep{ADNI_2010} for AD and the Parkinson's Progression Markers Initiative (PPMI)~\citep{PPMI_2011} for PD. Participants are eligible if they meet two criteria: (i) the availability of a baseline scan and at least one follow-up MRI, from which we select the first and last available scans, and (ii) the availability of the clinical variables required for downstream analysis. To maximize the sample size for each analysis, we define subsets of participants for each clinical variable based on data availability. For ADNI, the selected MRI visits are accompanied by Brain Parenchymal Fraction (BPF)~\citep{rudick1999use}, computed from the provided segmentations, as well as cognitive assessments aligned with the imaging time points, namely the Mini-Mental State Examination (MMSE)~\citep{cockrell2002mini}, the 13-item Alzheimer's Disease Assessment Scale--Cognitive Subscale (ADAS-13)~\citep{rosen1984new}, the Clinical Dementia Rating--Sum of Boxes (CDR-SB)~\citep{hughes1982new}, the Functional Activities Questionnaire (FAQ)~\citep{pfeffer1982measurement}, and the Montreal Cognitive Assessment (MoCA)~\citep{nasreddine2005montreal}. For PPMI, we select MRI visits paired with GM volume measurements, the Modified Schwab \& England Activities of Daily Living (MSEADLG) scale~\citep{schwab1969projection} and the MoCA. Table~\ref{tab:demographics} summarizes the demographic characteristics and reports the total number of patients included from each dataset, while Table~\ref{tab:adni_results} and Table~\ref{tab:ppmi_results} report the sample sizes associated with each clinical variable. Dataset statements can be found in Appendix~\ref{sec:dataset-statements}.

\begin{table}[pos=t] 
    \centering
    \caption{Summary of participants demographics across the evaluated cohorts.}
    \label{tab:demographics}
    \begin{tabular}{l cccc} 
        \toprule
        \textbf{Dataset} & $N$ & \textbf{Female (\%)} & \textbf{Age Mean} \\
        \midrule
        \textbf{ADNI} & 1006 & 445 (44\%) & 73.18 $\pm$ 7.07 \\
        \textbf{PPMI} & 310 & 133 (43\%) & 63.0 $\pm$ 8.90 \\
        \bottomrule
    \end{tabular}
\end{table}

\subsection{Correlation with disease progression}
To assess whether the proposed pipelines improve the clinical sensitivity of PBVC relative to \texttt{SIENA Vanilla}, we examine their associations with longitudinal changes in the cognitive, motor, and structural measures described in Section~\ref{sec:dataset}. Disease progression from baseline ($t_0$) to follow-up ($t_1$) is summarized by a Progression Index, $\Delta$, computed from the longitudinal change in each measure. Because some measures increase with worsening while others decrease, we define $\Delta$ for each variable such that $\Delta>0$ consistently indicates clinical worsening or structural atrophy. Table~\ref{tab:clinical_ref} reports the specific definition, including any sign reversal, used for each variable. Since PBVC is expected to be negative in our cohorts, reflecting volume loss, we correspondingly expect negative correlations between PBVC and all $\Delta$ measures. We quantify the linear association between each pipeline’s PBVC estimate and $\Delta$ using Pearson’s correlation coefficient ($r$). We summarize uncertainty in $r$ using 95\% confidence intervals (CIs) computed via Fisher’s $r$-to-$z$ transformation~\citep{fisher1915frequency}. To evaluate improvements in correlation, we conduct pairwise comparisons between each modified pipeline and \texttt{SIENA Vanilla} using Steiger’s $Z$ test for overlapping dependent correlations~\citep{steiger1980tests}. We assess statistical significance at $p < 0.01$, with Bonferroni correction to account for multiple comparisons.

\begin{table}[pos=t]
    \centering
    \caption{Standardization of the Progression Index ($\Delta$). Measures are grouped by the calculation required to ensure that a positive $\Delta$ consistently reflects disease progression, indicating either clinical decline or structural atrophy.}
    \label{tab:clinical_ref}
    \small
    \begin{tabularx}{\linewidth}{@{} l >{\raggedright\arraybackslash}X >{\raggedright\arraybackslash}X @{}}
        \toprule
        \textbf{$\Delta$ Calc.} & \textbf{Original Score} & \textbf{Included Measures} \\ 
        \midrule
        $t_0 - t_1$ & Decrease indicates worsening & MMSE, MoCA, GM Volume, BPF, MSEADLG \\
        \addlinespace
        $t_1 - t_0$ & Increase indicates worsening & ADAS-13, CDR-SB, FAQ \\
        \bottomrule
    \end{tabularx}
\end{table}

\subsection{Scan-Order consistency of PBVC estimates}
Ideally, SIENA should be invariant to the ordering of the input scans: exchanging the baseline and follow-up images should only change the sign of the estimated PBVC while preserving its magnitude. Formally, for a scan pair $(A,B)$, the estimated PBVC should satisfy
\begin{equation}
\text{PBVC}(A,B) = -\text{PBVC}(B,A).
\end{equation}
In practice, however, some steps of the pipeline may not be fully invariant to scan-order reversal, so the estimated PBVC may not exactly satisfy this antisymmetry property. To evaluate the robustness of each pipeline to this effect, we perform a scan-order consistency analysis. For every subject, the full pipeline is executed twice: first using the scans in their original temporal order $(A,B)$, and then after swapping the inputs $(B,A)$. We quantify the deviation from perfect antisymmetry using the absolute scan-order residual:
\begin{equation}
s(A,B) = \left| \text{PBVC}(A,B) + \text{PBVC}(B,A) \right|.
\end{equation}
A perfectly scan-order invariant pipeline would yield $s(A,B)=0$ for all scan pairs, up to numerical precision. Larger values indicate increasing sensitivity of the estimated atrophy to the ordering of the input scans.
To summarize this behavior across the dataset, we compute the Mean Forward--Reverse Residual (MFRR):
\begin{equation}\label{eq:3}
\text{MFRR} = \frac{1}{N} \sum_{i=1}^{N} s(A_i,B_i),
\end{equation}
where $N$ denotes the number of evaluated scan pairs. The MFRR is reported in absolute percentage points, along with its Standard Deviation (SD), and provides a measure of the numerical stability of the pipeline with respect to scan-order reversal. For ease of comparison, we also report the relative improvement of the proposed pipelines with respect to the \texttt{SIENA Vanilla} baseline
\begin{equation}\label{eq:4}
100 \times 
\frac{{\text{MFRR}}_{\text{vanilla}} - \text{MFRR}_{\text{pipe}}}
{\text{MFRR}_{\text{vanilla}}}.
\end{equation}
This analysis does not assess the biological accuracy of PBVC estimates. Rather, it serves as a measure of pipeline robustness.

\subsection{Execution time analysis}\label{subsec:exec_time_settings}
To evaluate computational efficiency and scalability, we measure the average end-to-end wall-clock execution time of each processing pipeline on 30 randomly selected subjects. Specifically, we benchmark the \texttt{SIENA Vanilla} baseline and the proposed variants under two execution configurations: (i) CPU-only processing and (ii) GPU-enabled processing for pipelines that support acceleration. For each subject pair, runtime is recorded for the full pipeline, from input ingestion to final output generation, and summary statistics (mean execution time in seconds) are computed across all evaluated pairs. All timings are obtained on the same workstation hardware and software environment described in Section~\ref{sec:impdetails}, ensuring a controlled comparison across methods.

\section{Results}\label{sec:results}
\subsection{Correlation between PBVC and disease progression in ADNI}
\begin{table*}[pos=t]
    \centering
    \caption{Comparative analysis of correlations between PBVC and longitudinal clinical changes in the ADNI cohort. The sample size ($N$) varies per clinical test and is denoted in the first column. The 95\% confidence intervals (CIs) were calculated using Fisher's $r$-to-$z$ transformation. Steiger's $Z$ test was utilized to compare the correlation coefficient of each modified pipeline against the \texttt{SIENA Vanilla} baseline.}
    \label{tab:adni_results}
    \footnotesize 
    \begin{tabular}{c l l c c c c}
        \toprule
        & \textbf{Test} & \textbf{Metric} & \textbf{\texttt{SIENA Vanilla}} & \textbf{\texttt{SIENA-SS}} & \textbf{\texttt{SIENA-SEG}} & \textbf{\texttt{SIENA-SS-SEG}} \\
        \midrule

        & $\Delta$MMSE & $r$ [95\% CI] & -0.226 [-0.284, -0.167] & \textbf{-0.497 [-0.542, -0.449]} & -0.252 [-0.309, -0.193] & -0.384 [-0.436, -0.331] \\
        & ($N=1006$)   & Steiger's $Z$ & \textit{Baseline}       & \textbf{-8.84} ($p < 0.001$)      & -1.14 ($p = 0.255$)      & -4.55 ($p < 0.001$) \\
        \cmidrule{2-7}

        & $\Delta$CDR-SB & $r$ [95\% CI] & -0.258 [-0.315, -0.199] & \textbf{-0.608 [-0.646, -0.567]} & -0.290 [-0.346, -0.232] & -0.453 [-0.501, -0.402] \\
        & ($N=985$)      & Steiger's $Z$ & \textit{Baseline}       & \textbf{-11.98} ($p < 0.001$)     & -1.41 ($p = 0.158$)      & -5.70 ($p < 0.001$) \\
        \cmidrule{2-7}

        & $\Delta$ADAS-13 & $r$ [95\% CI] & -0.254 [-0.311, -0.195] & \textbf{-0.524 [-0.567, -0.477]} & -0.271 [-0.328, -0.213] & -0.405 [-0.455, -0.352] \\
        & ($N=1006$)      & Steiger's $Z$ & \textit{Baseline}       & \textbf{-8.97} ($p < 0.001$)      & -0.75 ($p = 0.453$)      & -4.39 ($p < 0.001$) \\
        \cmidrule{2-7}

        & $\Delta$MoCA & $r$ [95\% CI] & -0.161 [-0.241, -0.079] & \textbf{-0.360 [-0.430, -0.285]} & -0.278 [-0.353, -0.199] & -0.357 [-0.428, -0.282] \\
        & ($N=553$)    & Steiger's $Z$ & \textit{Baseline}       & \textbf{-4.24} ($p < 0.001$)      & -4.01 ($p < 0.001$)      & -4.01 ($p < 0.001$) \\
        \cmidrule{2-7}

        & $\Delta$FAQ & $r$ [95\% CI] & -0.260 [-0.317, -0.201] & \textbf{-0.540 [-0.583, -0.494]} & -0.257 [-0.315, -0.198] & -0.394 [-0.446, -0.340] \\
        & ($N=981$)   & Steiger's $Z$ & \textit{Baseline}       & \textbf{-9.29} ($p < 0.001$)      & 0.11 ($p = 0.911$)       & -3.86 ($p < 0.001$) \\
        \cmidrule{2-7}

        & $\Delta$BPF & $r$ [95\% CI] & -0.118 [-0.179, -0.057] & \textbf{-0.249 [-0.306, -0.190]} & -0.098 [-0.159, -0.037] & -0.167 [-0.226, -0.106] \\
        & ($N=1006$)  & Steiger's $Z$ & \textit{Baseline}       & \textbf{-3.97} ($p < 0.001$)      & 0.84 ($p = 0.399$)       & -1.32 ($p = 0.186$) \\
        \bottomrule
    \end{tabular}
\end{table*}

Table~\ref{tab:adni_results} reports the correlation between PBVC and longitudinal changes in cognitive, functional, and structural measures in ADNI. Across all cognitive and functional indices, all pipelines produced the expected negative correlations (i.e., greater volume loss was associated with greater clinical worsening), with the strongest and most consistent improvements observed when replacing only the skull stripping module (\texttt{SIENA-SS}). Specifically, \texttt{SIENA-SS} increased the magnitude of correlation with $\Delta$MMSE from $r=-0.226$ to $r=-0.497$ ($Z=-8.84$, $p<0.001$), with $\Delta$ADAS-13 from $r=-0.254$ to $r=-0.524$ ($Z=-8.97$, $p<0.001$), and with $\Delta$FAQ from $r=-0.260$ to $r=-0.540$ ($Z=-9.29$, $p<0.001$). The largest gain was observed for $\Delta$CDR-SB, where \texttt{SIENA-SS} reached $r=-0.608$ compared to $r=-0.258$ for \texttt{SIENA Vanilla} ($Z=-11.98$, $p<0.001$), indicating substantially improved clinical sensitivity compared to \texttt{SIENA Vanilla}. Replacing only the segmentation step (\texttt{SIENA-SEG}) yielded comparatively modest changes and did not significantly differ from \texttt{SIENA Vanilla} for most measures, with the notable exception of $\Delta$MoCA where \texttt{SIENA-SEG} improved the association from $r=-0.161$ to $r=-0.278$ ($Z=-4.01$, $p<0.001$). The fully integrated pipeline (\texttt{SIENA-SS-SEG}) consistently improved correlations relative to \texttt{SIENA Vanilla} (e.g., $\Delta$CDR-SB: $r=-0.453$, $Z=-5.70$, $p<0.001$), but typically remained below \texttt{SIENA-SS} alone for clinical association strength. Finally, correlations with the structural reference $\Delta$BPF were weaker overall, yet \texttt{SIENA-SS} still significantly improved the association with respect to \texttt{SIENA Vanilla} (from $r=-0.118$ to $r=-0.249$, $Z=-3.97$, $p<0.001$).

\subsection{Correlation between PBVC and disease progression in PPMI}

\begin{table*}[pos=t]
    \centering
    \caption{Comparative analysis of correlations between PBVC and longitudinal clinical changes in the PPMI cohort. The sample size ($N$) varies per clinical test and is denoted in the first column. The 95\% confidence intervals (CIs) were calculated using Fisher's $r$-to-$z$ transformation. Steiger's $Z$ test was utilized to compare the correlation coefficient of each modified pipeline against the \texttt{SIENA Vanilla} baseline, with $p$-values Bonferroni-corrected for three comparisons per test.}
    \label{tab:ppmi_results}
    \footnotesize 
    \begin{tabular}{c l l c c c c}
        \toprule
        & \textbf{Test} & \textbf{Metric} & \textbf{\texttt{SIENA Vanilla}} & \textbf{\texttt{SIENA-SS}} & \textbf{\texttt{SIENA-SEG}} & \textbf{\texttt{SIENA-SS-SEG}} \\
        \midrule

        & $\Delta$MSEADLG & $r$ [95\% CI] & -0.058 [-0.171, 0.057] & \textbf{-0.126 [-0.238, -0.012]} & -0.041 [-0.156, 0.074] & -0.116 [-0.228, -0.001] \\
        & ($N=291$)       & Steiger's $Z$ & \textit{Baseline}      & -0.64 ($p = 0.520$)      & \textbf{0.93} ($p = 0.352$)      & -0.41 ($p = 0.681$) \\
        \cmidrule{2-7}

        & $\Delta$MoCA & $r$ [95\% CI] & 0.021 [-0.163, 0.205] & \textbf{-0.122 [-0.299, 0.064]} & -0.066 [-0.247, 0.119] & -0.117 [-0.294, 0.069] \\
        & ($N=114$)    & Steiger's $Z$ & \textit{Baseline}     & \textbf{-1.73} ($p = 0.253$)     & -0.84 ($p = 1.000$)    & -1.54 ($p = 0.368$) \\
        \cmidrule{2-7}

        & $\Delta$GM & $r$ [95\% CI] & -0.020 [-0.199, 0.160] & \textbf{-0.219 [-0.384, -0.042]} & -0.151 [-0.321, 0.029] & -0.216 [-0.381, -0.038] \\
        & ($N=120$)  & Steiger's $Z$ & \textit{Baseline}      & \textbf{-2.19} ($p = 0.087$)      & -1.21 ($p = 0.677$)     & -2.05 ($p = 0.120$) \\
        \bottomrule
    \end{tabular}
\end{table*}
Table~\ref{tab:ppmi_results} summarizes the correlations between PBVC and longitudinal clinical changes in the PPMI cohort. Overall effect sizes were small in this subset, and the \texttt{SIENA Vanilla} baseline showed near-null associations across all outcomes ($r=0.021$ for $\Delta$MoCA; $r=-0.058$ for $\Delta$MSEADLG; $r=-0.020$ for $\Delta$GM). For $\Delta$MoCA, \texttt{SIENA-SS} yielded the strongest negative association ($r=-0.122$), with \texttt{SIENA-SEG} ($r=-0.066$) and \texttt{SIENA-SS-SEG} ($r=-0.117$) showing similar directional behavior; however, none of these differences versus baseline were significant by Steiger’s test after Bonferroni correction (\texttt{SIENA-SS}: $Z=-1.73$, $p=0.253$). A similar pattern was observed for $\Delta$MSEADLG, where the modified pipelines shifted the relationships toward the expected negative direction when skull stripping was replaced, with \texttt{SIENA-SS} yielding the strongest negative association ($r=-0.126$) and \texttt{SIENA-SS-SEG} showing a similar effect ($r=-0.116$), whereas \texttt{SIENA-SEG} remained close to baseline ($r=-0.041$); again, none of these differences versus baseline were significant after correction (\texttt{SIENA-SS}: $Z=-0.64$, $p=0.520$). The clearest gain was observed for $\Delta$GM, where \texttt{SIENA-SS} increased the magnitude of the negative correlation to $r=-0.219$ (\texttt{SIENA-SS-SEG}: $r=-0.216$; \texttt{SIENA-SEG}: $r=-0.151$), again without reaching corrected significance (\texttt{SIENA-SS}: $Z=-2.19$, $p=0.087$).

\begin{figure*}[pos=t]
     \centering
     \includegraphics[width=\linewidth]{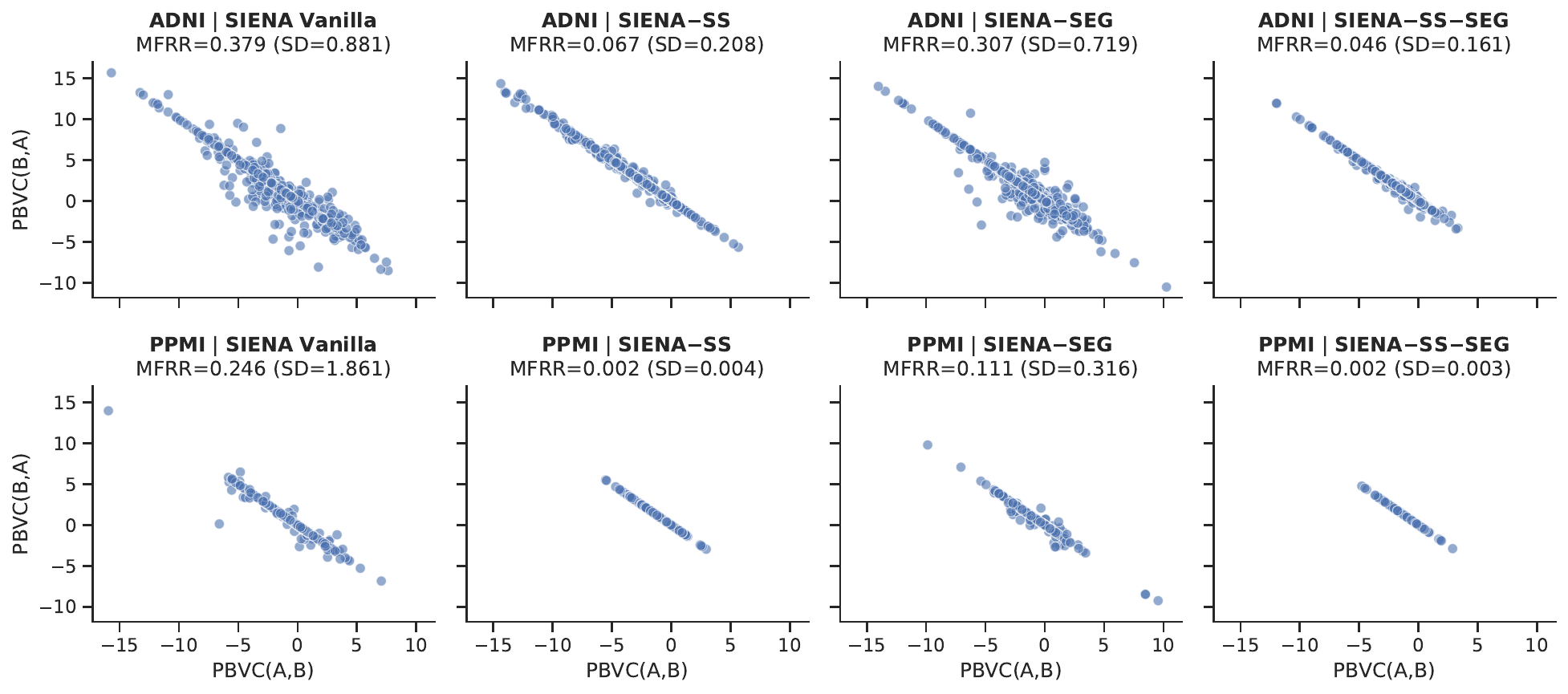}
     \caption{Scan-order consistency of PBVC estimates across pipelines and datasets. Scatter plots compare the estimated PBVC for the original temporal scan order, $\mathrm{PBVC}(A,B)$ (x-axis), with that for the reversed scan order, $\mathrm{PBVC}(B,A)$ (y-axis), in the ADNI (top row) and PPMI (bottom row) datasets. A perfectly scan-order-invariant pipeline would yield points lying exactly on the antisymmetry diagonal. The MFRR and its standard deviation are reported for each configuration.}
     \label{fig:scan-order}
\end{figure*}

\subsection{Scan-order consistency results}\label{sec:simmetry}
Figure~\ref{fig:scan-order} summarizes the scan-order consistency results across the evaluated datasets and pipelines. Across both datasets, the \texttt{SIENA Vanilla} baseline exhibited the highest error, with an MFRR of 0.379\% (SD 0.881\%) in ADNI ($N=1006$) and 0.246\% (SD 1.861\%) in PPMI ($N=310$). Relative to this baseline, all modified pipelines reduced MFRR. In ADNI, \texttt{SIENA-SEG} achieved an 18.9\% reduction (MFRR 0.307\%, SD 0.719\%), while \texttt{SIENA-SS} yielded a substantially larger 82.4\% reduction (MFRR 0.067\%, SD 0.208\%). The combined \texttt{SIENA-SS-SEG} pipeline provided the greatest improvement, reducing the residual by 87.8\% and attaining the lowest MFRR of 0.046\% (SD 0.161\%). In PPMI, \texttt{SIENA-SEG} reduced MFRR by 54.8\% (MFRR 0.111\%, SD 0.316\%), whereas both \texttt{SIENA-SS} and \texttt{SIENA-SS-SEG} showed near-complete error reduction relative to baseline. Specifically, \texttt{SIENA-SS} achieved a 99.0\% reduction, corresponding to an MFRR of 0.002\% (SD 0.004\%), while \texttt{SIENA-SS-SEG} achieved a similar 99.1\% reduction with the same MFRR of 0.002\% and a slightly lower variability (SD 0.003\%).

\subsection{Execution time results}
Table~\ref{tab:time_executions} reports the average end-to-end runtime of \texttt{SIENA Vanilla} and the proposed variants. On CPU, \texttt{SIENA Vanilla} required an average of 1855.1\,s per subject pair (approximately 31\,min). Under the same CPU-only setting, \texttt{SIENA-SS}, \texttt{SIENA-SEG}, and \texttt{SIENA-SS-SEG} achieved comparable runtimes (ranging from 1827.4 s to 1851.0 s), indicating that the introduced modules do not increase overall processing time. When GPU acceleration was enabled, the proposed pipelines achieved clear speedups. \texttt{SIENA-SS} showed a modest reduction (1790.8\,s), whereas methods incorporating the tissue segmentation module exhibited the largest gains: \texttt{SIENA-SEG} and \texttt{SIENA-SS-SEG} reduced mean runtime to 1001.9\,s (approximately 16.7\,min) and 1166.4\,s (approximately 19.4\,min), respectively.

\begin{table}[pos=t]
    \centering
    \caption{End-to-end execution time (in seconds) for each pipeline. The results compare CPU and GPU processing performance between the four examined pipelines.}
    \small
    \begin{tabular}{ll cc}
        \toprule
        \textbf{Pipeline} & \textbf{Device} & \textbf{Mean (s)} & \textbf{SD (s)} \\
        \midrule
        \multirow{2}{*}{\texttt{SIENA Vanilla}} 
        & CPU & 1855.131 & 304.985 \\
        & GPU & -- & -- \\
        \midrule
        \multirow{2}{*}{\texttt{SIENA-SS}} 
        & CPU & 1827.353 & 194.975 \\
        & GPU & 1790.759 & 413.627 \\
        \midrule
        \multirow{2}{*}{\texttt{SIENA-SEG}} 
        & CPU & 1851.025 & 293.543 \\
        & GPU & \textbf{1001.921} & 153.251\\
        \midrule
        \multirow{2}{*}{\texttt{SIENA-SS-SEG}} 
        & CPU & 1849.442 & 228.318 \\
        & GPU & 1166.433 & 350.194 \\
        \bottomrule
    \end{tabular}
    \label{tab:time_executions}
\end{table}

\section{Discussion}\label{sec:discussion}
The objective of this study is to determine whether targeted DL substitutions in the image processing stages of SIENA improve the clinical sensitivity and robustness of longitudinal brain atrophy estimates while preserving the core methodology of the original pipeline. To this end, we replace SIENA's classical skull-stripping and tissue-segmentation modules with SynthStrip and SynthSeg, respectively, and evaluate the resulting pipeline variants in ADNI and PPMI.

The results show that updating SIENA's image processing steps increases the association between PBVC and longitudinal measures of disease progression, with the most consistent gains arising from the replacement of the skull-stripping step. In ADNI, \texttt{SIENA-SS} shows substantially stronger correlations with clinical worsening than \texttt{SIENA Vanilla} across all evaluated measures. By contrast, replacing only the segmentation module yields smaller and less consistent changes. The fully integrated pipeline, \texttt{SIENA-SS-SEG}, generally improves on the vanilla baseline but often does not outperform \texttt{SIENA-SS} alone in terms of correlation strength. One possible explanation is that the two methods capture different aspects of brain anatomy. Whereas SynthSeg~\citep{billot2023synthseg} is trained to segment anatomical structures, FAST~\citep{Fast} directly models tissue classes. For a framework such as SIENA, which depends on accurate tissue boundary estimation, this difference in emphasis may influence downstream performance.

In PPMI, the proposed pipelines consistently shifted the associations with the analyzed progression measures in the expected direction, particularly in the variants replacing the skull-stripping step, further supporting the idea that this component is the main driver of improvement in the proposed framework. At the same time, these effects were modest and did not reach statistical significance relative to the standard SIENA baseline after multiple-comparison correction, highlighting the need for further validation in additional cohorts to better define the extent of the benefit in this setting.

The scan-order consistency analysis highlights that all modified pipelines reduce the MFRR relative to \texttt{SIENA Vanilla}. In particular, the integrated \texttt{SIENA-SS-SEG} variant achieved the lowest MFRR across both datasets. This represents a marked improvement over \texttt{SIENA Vanilla} (87.8\% in ADNI and 99.1\% in PPMI) indicating superior robustness to scan-order reversal. In contrast to the correlation experiment, replacing FAST with SynthSeg proves beneficial. This result may reflect the design of SynthSeg, which was trained on highly heterogeneous synthetic data with the aim of improving robustness~\citep{billot2023synthseg}.

The execution-time analysis adds a practical dimension to these results: on CPU, the proposed pipelines achieve end-to-end runtimes comparable to \texttt{SIENA Vanilla}. With GPU acceleration, substituting FAST with SynthSeg decreases processing time by as much as 14 minutes compared with the other variants. In contrast, replacing SynthStrip yields only a modest improvement, since BET2 already has a low runtime.

Several limitations of our study should be noted. First, since we chose to focus on the SIENA framework due to its widespread adoption, we did not consider alternative brain atrophy measures. We leave a comprehensive cross-framework comparison for future work. Second, because obtaining in vivo ground truth for brain atrophy is extremely difficult~\citep{de2011comparison}, the assessment of the proposed methods relied on surrogate endpoints, such as the relationship between PBVC and clinical indicators of disease progression.

These findings point to several promising directions for future work. First, developing or identifying validation frameworks that more closely approximate ground truth for neurodegeneration could enable a more direct assessment of PBVC accuracy and reduce reliance on indirect evaluation criteria. For example, generative models of brain changes~\citep{litrico2024tadm,puglisi2025brain} could potentially serve this purpose, although they have not yet reached a sufficient level of control over the simulated patterns. Second, the pipeline may benefit from broader use of DL, for example by replacing additional processing bottlenecks, which could enable faster brain atrophy estimation and facilitate large-scale application.

\section{Conclusions}
In this work, we investigated whether a DL-based modernization of the SIENA pipeline could improve longitudinal brain atrophy estimation while preserving SIENA’s established core methodology. To this end, classical image processing components were selectively replaced with the DL tools SynthStrip and SynthSeg within an otherwise unchanged framework. Across two large neurodegenerative cohorts, the results showed that improving image processing robustness, particularly at the skull-stripping stage, can strengthen the clinical relevance of PBVC estimates, improve scan-order consistency, and, when GPU acceleration is available, reduce end-to-end computation time. These findings indicate that the targeted integration of modern DL components into interpretable and clinically established pipelines can yield meaningful gains in sensitivity, robustness, and scalability without sacrificing methodological transparency. More broadly, this study supports a modular strategy for modernizing legacy neuroimaging tools, in which DL is used not to replace validated analytical frameworks, but to reinforce their weakest processing stages and thereby improve their utility for large-scale longitudinal studies and clinical research.


\section*{Data availability}
The link to the code is provided in the manuscript. Access to the ADNI and PPMI datasets can be requested through the LONI IDA portal (\url{https://ida.loni.usc.edu/}).

\section*{Acknowledgments}

\textit{Data collection and sharing for the Alzheimer's Disease Neuroimaging Initiative (ADNI) is funded by the National Institute on Aging (National Institutes of Health Grant U19AG024904). The grantee organization is the Northern California Institute for Research and Education. In the past, ADNI has also received funding from the National Institute of Biomedical Imaging and Bioengineering, the Canadian Institutes of Health Research, and private sector contributions through the Foundation for the National Institutes of  Health (FNIH) including generous contributions from the following: AbbVie, Alzheimer’s Association; Alzheimer’s Drug Discovery Foundation; Araclon Biotech; BioClinica, Inc.; Biogen; Bristol-Myers Squibb Company; CereSpir, Inc.; Cogstate; Eisai Inc.; Elan Pharmaceuticals, Inc.; Eli Lilly and Company; EuroImmun; F. Hoffmann-La Roche Ltd and its affiliated company Genentech, Inc.; Fujirebio; GE Healthcare; IXICO Ltd.; Janssen Alzheimer Immunotherapy Research \& Development, LLC.; Johnson \& Johnson Pharmaceutical Research \& Development LLC.; Lumosity; Lundbeck; Merck \& Co., Inc.; Meso Scale Diagnostics, LLC.; NeuroRx Research; Neurotrack Technologies; Novartis Pharmaceuticals Corporation; Pfizer Inc.; Piramal Imaging; Servier; Takeda Pharmaceutical Company; and Transition Therapeutics} \newline

\textit{Data used in the preparation of this article was obtained on 2025-05-20 from the Parkinson’s
Progression Markers Initiative (PPMI) database (\url{www.ppmi-info.org/access-data-specimens/download-data}), RRID: SCR\_006431. For up-to-date information on the study, visit \url{www.ppmi-info.org.} 
PPMI – a public-private partnership – is funded by the Michael J. Fox Foundation for Parkinson’s Research and funding partners (full list available at \url{https://www.ppmi-info.org/about-ppmi/who-we-are/study-sponsors}).}

\printcredits
\bibliographystyle{cas-model2-names}
\bibliography{cas-refs}

\appendix

\section{Dataset statements}\label{sec:dataset-statements}
\textit{Data used in the preparation of this article were obtained from the Alzheimer's Disease Neuroimaging Initiative (ADNI) database (\url{adni.loni.usc.edu}). The ADNI was launched in 2003 as a public-private partnership, led by Principal Investigator Michael W. Weiner, MD. The original goal of ADNI was to test whether serial MRI, positron emission tomography (PET), other biological markers, and clinical and neuropsychological assessment can be combined to measure the progression of MCI and early AD. The current goals include validating biomarkers for clinical trials, improving the generalizability of ADNI data by increasing diversity in the participant cohort, and to provide data concerning the diagnosis and progression of AD to the scientific community. For up-to-date information, see \url{adni.loni.usc.edu}.}

\end{document}